\documentclass[showpacs,twocolumn,preprintnumbers,amsmath,amssymb]{revtex4}

\usepackage{graphicx}
\usepackage{bm}
\usepackage{color}

\newcommand{\bq}{\begin{equation}}
\newcommand{\eq}{\end{equation}}
\newcommand{\bqa}{\begin{eqnarray}}
\newcommand{\eqa}{\end{eqnarray}}
\newcommand{\nn}{\nonumber \\}

\def\be     {\begin{equation}}
\def\ee     {\end{equation}}
\def\bea        {\begin{eqnarray}}
\def\eea        {\end{eqnarray}}
\def\bnn    {\begin{eqnarray*}}
\def\enn    {\end{eqnarray*}}

\begin{document}

\title{An emergent geometric description for a topological phase transition in the Kitaev superconductor model}
\author{Ki-Seok Kim$^{1}$, Miok Park$^{2,3}$, Jaeyoon Cho$^{1,3}$, and Chanyong Park$^{1,3}$}
\affiliation{$^{1}$Department of Physics, POSTECH, Pohang, Gyeongbuk 790-784, Korea \\ $^{2}$Center for Artificial Low Dimensional Electronic Systems, Institute for Basic Science (IBS), 77 Cheongam-Ro, Pohang, Gyeongbuk 790-784, Korea \\ $^{3}$Asia Pacific Center for Theoretical Physics (APCTP), POSTECH, Pohang, Gyeongbuk 790-784, Korea}

\date{\today}

\begin{abstract}
Resorting to Wilsonian renormalization group (RG) transformations, we propose an emergent geometric description for a topological phase transition in the Kitaev superconductor model. An effective field theory consists of an emergent bulk action with an extra dimension, an ultraviolet (UV) boundary condition for an initial value of a coupling function, and an infrared (IR) effective action with a fully renormalized coupling function. The bulk action describes the evolution of the coupling function along the direction of the extra dimension, where the extra dimension is identified with an RG scale and the resulting equation of motion is nothing but a $\beta-$function. In particular, the IR effective field theory turns out to be consistent with a Callan-Symanzik equation which takes into account both the bulk and IR boundary contributions. This derived Callan-Symanzik equation gives rise to a metric structure. Based on this emergent metric tensor, we uncover the equivalence of the entanglement entropy between the emergent geometric description and the quantum field theory in the vicinity of the quantum critical point.
\end{abstract}


\maketitle

\section{Introduction}

Recent experiments proposed that hydrodynamics may be realized in Dirac band systems at intermediate temperatures \cite{Emergent_Hydrodynamics_Experiment1,Emergent_Hydrodynamics_Experiment2,Emergent_Hydrodynamics_Experiment3}, where incomplete screening of electron correlations due to the band structure can allow the time scale of electron-electron scattering being shorter than the others of electron-phonon and electron-impurity scattering time scales. AdS$_{d+2}$/CFT$_{d+1}$ duality conjecture \cite{AdS_CFT_Original,AdS_CFT_Follow,AdS_CFT_Correspondence,AdS_CFT_Review} looks well devised to describe such emergent hydrodynamics \cite{AdS_Hydrodynamics}, where fast thermalization due to strong inelastic scattering may play a central role. Recent theoretical calculations based on the AdS$_{4}$/CFT$_{3}$ duality conjecture could fit the experimental data for both electrical and thermal transport coefficients in graphene-type band structures surprisingly well \cite{Graphene_AdS_Description1,Graphene_AdS_Description2}. Unfortunately, these kinds of successful examples for the comparison between the AdS$_{d+2}$/CFT$_{d+1}$ duality conjecture and actual experimental data would have some limits even if this comparison may not be allowed within the perturbative theoretical framework. An essential point behind this situation is that we do not know how calculations based on the AdS$_{d+2}$/CFT$_{d+1}$ duality conjecture can be related with microscopic degrees of freedom.

Recent developments in the derivation of the AdS$_{d+2}$/CFT$_{d+1}$ duality conjecture from field theoretical perspectives \cite{RG_AdS_CFT1,RG_AdS_CFT2,RG_AdS_CFT3,RG_AdS_CFT4,RG_AdS_CFT5,RG_AdS_CFT6,RG_AdS_CFT7,RG_AdS_CFT8,RG_AdS_CFT9,RG_AdS_CFT10,RG_AdS_CFT11,RG_AdS_CFT12,From_Field_Theory_To_Emergent_Gravity,Emergence_Extra_Dimension} are based on either how to implement the Wilson's renormalization group (RG) structure \cite{Wilson_RG_Review} to field theory or how to reformulate quantum information (entanglement entropy or complexity) in terms of geometry. In particular, the multi-scale entanglement renormalization ansatz (MERA) has been proposed to represent a many-particle ground-state wave-function, applying the Wilson's RG transformation in a non-perturbative way \cite{MERA1,MERA2,MERA_Review}. The number of RG transformations is identified with an emergent extra dimension \cite{Swingle_MERA_AdS}. Comparing the entanglement entropy \cite{Entanglement_Entropy_Review} based on the MERA representation with the Ryu-Takayanagi formula \cite{Ryu_Takayanagi_Formula} based on the AdS$_{d+2}$/CFT$_{d+1}$ duality conjecture, the metric structure of the emergent AdS$_{d+2}$ space has been discussed \cite{MERA_AdS_CFT}. Recently, the partition function in the path-integral representation has been reformulated based on the MERA construction, realizing an emergent holographic description with an extra dimension \cite{From_Field_Theory_To_Emergent_Gravity}. In particular, it turns out that this MERA-based gravity reformulation for a field theory coincides with a previous path-integral reformulation for the implementation of the quantum RG construction.

Applying Wilsonian RG transformations into the Kitaev superconductor model, we find an effective field theory with an extra dimension: (1) Two auxiliary fields are introduced to play the role of a coupling function and an order parameter field, respectively, where they form a canonical conjugate pair. (2) A bulk action describes the evolution of the coupling function along the direction of the emergent extra dimension, where the extra dimension is identified with an RG scale and the resulting equation of motion is nothing but a $\beta-$function. (3) An ultraviolet (UV) boundary condition defines an initial value of the coupling function. (4) An infrared (IR) boundary condition is given by an effective theory with a fully renormalized coupling function, which evolves through the extra dimension. In particular, the IR effective field theory turns out to be consistent with a Callan-Symanzik equation for an effective free energy which takes into account both the bulk and IR boundary contributions. Based on this derived Callan-Symanzik equation, we extract out a metric structure. In order to justify this geometric description, we calculate the holographic entanglement entropy based on the Ryu-Takayanagi formula and compare it with the entanglement entropy of the corresponding field theory for the whole parameter range of the topological phase transition. We reveal the equivalence of the entanglement entropy between the emergent geometric description and the quantum field theory not only at the quantum critical point but also in the topological superconductor phase.

\section{Kadanoff block-spin transformation and emergence of an extra dimension}

We introduce the transverse-field Ising model
\bqa && H = - \frac{J}{2} \sum_{i = 1}^{N} \Big( \sigma_{i}^{z} \sigma_{i+1}^{z} + \lambda \sigma_{i}^{x} \Big) . \eqa
Here, $J$ is a ferromagnetic coupling constant, and $\lambda$ is a transverse magnetic field. This Hamiltonian enjoys Z$_{2}$ symmetry, given by $\sigma_{i}^{z} \rightarrow - \sigma_{i}^{z}$. It is well known that this model shows a quantum phase transition at zero temperature from a ferromagnetic phase of $\langle \sigma_{i}^{z} \rangle \not= 0$ in the case of $\lambda < \lambda_{c}$ to a paramagnetic state of $\langle \sigma_{i}^{z} \rangle = 0$ in the case of $\lambda > \lambda_{c}$ \cite{QFT_One_Dimension}. This continuous phase transition becomes smeared out at finite temperatures due to low dimensionality.

Recently, this Z$_{2}$ symmetry breaking transition has been revisited in the novel perspective of a topological phase transition \cite{Kitaev_Superconductor_Model}. We introduce the Jordan-Wigner transformation $\sigma_{i}^{x} = 2 c_{i}^{\dagger} c_{i} - 1$ and $\sigma_{i}^{z} = (-1)^{i-1} e^{\pm i \pi \sum_{j = 1}^{i-1} c_{j}^{\dagger} c_{j}} (c_{i}^{\dagger} + c_{i})$ \cite{QFT_One_Dimension}, where $c_{i}$ is a spinless fermion field. Physically speaking, these spinless fermions describe domain wall excitations. Then, we map the transverse-field Ising model into a superconductor model in terms of spinless fermions, referred to as a Kitaev model,
\bqa && Z = \int \Pi_{i = 1}^{N} D \psi_{i} \exp\Bigl[ - \int_{0}^{\beta} d \tau \sum_{i = 1}^{N} \Bigl\{ \psi_{i}^{\dagger} \Big( \partial_{\tau} I + J \lambda \tau_{3} \Big) \psi_{i} \nn && - J \psi_{i}^{\dagger} \Big( \tau_{3} - i \tau_{2} \Big) \psi_{i+1} \Bigr\} \Bigr] , \label{Kitaev_Model} \eqa
where the original expression of the Kitaev model has been reformulated in terms of the Nambu-spinor representation $\psi_{i} = \left( \begin{array} [c]{c} c_{i} \\ c_{i}^{\dagger} \end{array} \right)$ \cite{BCS_Textbook} with rescaling of the fermion field. $I$ is an identity matrix and $\tau_{i}$ with $i = 1, 2, 3$ is a Pauli matrix. It turns out that this superconductor model shows a phase transition from a p-wave weak-pairing (BCS) superconducting phase to a p-wave strong-pairing (BEC) superconducting state, tuning the chemical potential $\lambda$ at zero temperature \cite{Kitaev_Superconductor_Model}. The BCS superconducting phase of $\lambda < \lambda_{c}$ is identified with a topological superconducting state, where there exists a Majorana zero-energy state at each edge. On the other hand, the BEC superconducting phase of $\lambda > \lambda_{c}$ is identified with a normal superconducting state without any edge states. The bulk gap becomes closed at the quantum critical point of $\lambda = \lambda_{c}$.

In order to construct an emergent geometric description for this topological phase transition, we introduce collective fields as follows
%
%
\bqa && Z = \int D \chi^{(0)} D \eta^{(0)} \Pi_{i = 1}^{N} D \psi_{i} \nn && \exp\Bigl[ - \sum_{i \omega} \sum_{i = 1}^{N} \Bigl\{ \psi_{i}^{\dagger} \Big( - i \omega I + J \lambda \tau_{3} \Big) \psi_{i} \nn && - \chi^{(0)} \psi_{i}^{\dagger} ( \tau_{3} - i \tau_{2} ) \psi_{i+1} + \eta^{(0)} (\chi^{(0)} - J) \Bigr\} \Bigr] , \label{Kitaev_Model_Collective_Fields} \eqa
where $\int D \chi^{(0)} D \eta^{(0)}$ in this expression reproduces Eq. (\ref{Kitaev_Model}). $\chi^{(0)}$ is a Lagrange multiplier field to play the role of a coupling function and $\eta^{(0)} = \Big\langle \psi_{i}^{\dagger} ( \tau_{3} - i \tau_{2} ) \psi_{i+1} \Big\rangle$ is an order-parameter field in the saddle-point analysis. It turns out that they form a canonical conjugate pair.

Now, we apply the Wilson's RG transformation. First, we separate the site index into even and odd. Second, we perform the gaussian integration for odd-site fermion fields. Third, we take into account rescaling for both the lattice structure and the fermion field, reproducing the original form of the partition function. As a result, we obtain
\bqa && Z = \int D \chi^{(0)} D \eta^{(0)} \Pi_{i = 1}^{N} D \psi_{i} \nn && \exp\Bigl[ - \sum_{i \omega} \sum_{i = 1}^{N} \Bigl\{ \psi_{i}^{\dagger} \Big( - i \omega I + J \lambda \tau_{3} \Big) \psi_{i} \nn && - f(\chi^{(0)}) \psi_{i}^{\dagger} ( \tau_{3} - i \tau_{2} ) \psi_{i+1} + \eta^{(0)} (\chi^{(0)} - J) \Bigr\} \Bigr] , \label{Kitaev_Model_Collective_Fields_RG_First_Iteration} \eqa
where the coupling function becomes renormalized from $\chi^{(0)}$ to $f(\chi^{(0)}) = \frac{2 J \lambda}{\omega^{2} + (J \lambda)^{2}} [\chi^{(0)}]^{2}$. Once again, we introduce collective fields and rewrite Eq. (\ref{Kitaev_Model_Collective_Fields_RG_First_Iteration}) in the following way
\bqa && Z = \int D \chi^{(0)} D \eta^{(0)} D \chi^{(1)} D \eta^{(1)} \Pi_{i = 1}^{N} D \psi_{i} \nn && \exp\Bigl[ \sum_{i \omega} \sum_{i = 1}^{N} \Bigl\{ \psi_{i}^{\dagger} \Big( - i \omega I + J \lambda \tau_{3} \Big) \psi_{i} - \chi^{(1)}  \psi_{i}^{\dagger} ( \tau_{3} - i \tau_{2} ) \psi_{i+1} \nn && + \eta^{(0)} (\chi^{(0)} - J) + \eta^{(1)} \Big( \chi^{(1)} - f(\chi^{(0)}) \Big) \Bigr\} \Bigr] , \eqa
where $\int D \chi^{(1)} D \eta^{(1)}$ gives rise to Eq. (\ref{Kitaev_Model_Collective_Fields_RG_First_Iteration}).

Repeating these RG transformations, we obtain
\bqa && Z = \int D \chi^{(0)} D \eta^{(0)} \Pi_{k = 1}^{f} D \chi^{(k)} D \eta^{(k)} \Pi_{i = 1}^{N} D \psi_{i} \nn && \exp\Bigl[ - \sum_{i \omega} \sum_{i = 1}^{N} \Bigl\{ \psi_{i}^{\dagger} \Big( - i \omega I + J \lambda \tau_{3} \Big) \psi_{i} \nn && - \chi^{(f)} \psi_{i}^{\dagger} ( \tau_{3} - i \tau_{2} ) \psi_{i+1} + \eta^{(0)} (\chi^{(0)} - J) \Big\} \nn && - N \sum_{k = 1}^{f} \sum_{i \omega} \eta^{(k)} \Big( \chi^{(k)} - f(\chi^{(k-1)}) \Big) \Bigr] , \label{Holography_Discrete} \eqa
where $f(\chi^{(k-1)}) = \frac{2 J \lambda}{\omega^{2} + (J \lambda)^{2}} [\chi^{(k-1)}]^{2}$ is a renormalized coupling function after the $k-$th iteration.

An idea is to translate the iteration index $k$ of RG transformations into an extra coordinate $z$ as follows
\begin{widetext}
\bqa && Z = \int \Pi_{i = 1}^{N} D \psi_{i}(i\omega) D \chi(i\omega,z) D \eta(i\omega,z) \exp\Big\{-  \mathcal{S}_{UV}[\eta(i \omega,0), \chi(i \omega,0)] - \mathcal{S}_{Bulk}[\eta(i \omega,z), \chi(i \omega,z)] \nn && - \mathcal{S}_{IR}[\psi_{i}(i\omega);\chi(i \omega,z_{f})] \Big\} ,  \label{Kitaev_Model_Emergent_Holography} \\ && \mathcal{S}_{UV}[\eta(i \omega,0), \chi(i \omega,0)] = N \eta(i \omega,0) \Big( \chi(i \omega,0) - J \Big) , \\ && \mathcal{S}_{Bulk}[\eta(i \omega,z), \chi(i \omega,z)] = N \int_{0}^{z_{f}} d z \sum_{i \omega} \eta(i \omega,z) \Big( \frac{\partial \chi(i \omega,z) }{\partial z} + \chi(i \omega,z) - f[\chi(i \omega,z)] \Big) , \\ &&
\mathcal{S}_{IR}[\psi_{i}(i\omega);\chi(i \omega,z_{f})] = \sum_{i \omega} \sum_{i = 1}^{N} \Bigl\{ \psi_{i}^{\dagger}(i \omega) \Big( - i \omega I + J \lambda \tau_{3} \Big) \psi_{i}(i \omega) - \chi(i \omega,z_{f}) \psi_{i}^{\dagger}(i \omega) ( \tau_{3} - i \tau_{2} ) \psi_{i+1}(i \omega) \Big\} . \eqa
\end{widetext}
%
%
$\mathcal{S}_{Bulk}[\eta(i \omega,z), \chi(i \omega,z)]$ is an effective bulk action, where $\eta(i \omega,z) \frac{\partial \chi(i \omega,z) }{\partial z}$ implies that $\eta(i \omega,z)$ and $\chi(i \omega,z)$ are a canonical conjugate pair in the Hamiltonian formulation. It gives rise to the $\beta_{\chi} \equiv \frac{\partial \chi(i \omega,z) }{\partial z}$ function
%
%
\bqa && \frac{\partial \chi(i \omega,z) }{\partial z} = - \chi(i \omega,z) + f[\chi(i \omega,z)] , \label{Beta_Function} \eqa
where
\bqa && f[\chi(i \omega,z)] = \frac{2 J \lambda}{\omega^{2} + (J \lambda)^{2}} [\chi(i \omega,z)]^{2} \eqa
is a renormalized coupling function, resulting from RG transformations. $\mathcal{S}_{UV}[\eta(i \omega,0), \chi(i \omega,0)]$ describes a UV boundary condition of the coupling function, resulting in $\chi(i \omega,0) = J$. This coupling function evolves into $\chi(i \omega,z_{f})$ according to the $\beta_{\chi}$ function. It is almost trivial to solve Eq. (\ref{Beta_Function}), giving rise to
\bqa && \chi(i \omega,z_{f}) = \frac{ \Big\{1 + \Big(\frac{\omega}{J\lambda}\Big)^{2}\Big\} J \lambda}{2 + \Big\{\lambda - 2 + \lambda \Big(\frac{\omega}{J\lambda}\Big)^{2}\Big\} e^{z_{f}}} . \eqa
As a result, we obtain an effective IR action $\mathcal{S}_{IR}[\psi_{i}(i\omega);\chi(i \omega,z_{f})]$ with a fully renormalized coupling function $\chi(i \omega,z_{f})$.

In order to complete our construction of an emergent geometric description, we integrate over original fermion degrees of freedom. Performing the Fourier transformation and the Gaussian integration for the spinor field, we obtain an effective action as follows
\begin{widetext}
\bqa && Z = \int D \chi(i\omega,z) D \eta(i\omega,z) \exp\Big\{- \mathcal{S}_{UV}[\eta(i \omega,0), \chi(i \omega,0)] - \mathcal{S}_{Bulk}[\eta(i \omega,z), \chi(i \omega,z)] - \mathcal{S}_{IR}[\chi(i \omega,z_{f})] \Big\} ,  \label{Kitaev_Model_Emergent_Holography_Bosonic} \\ && \mathcal{S}_{UV}[\eta(i \omega,0), \chi(i \omega,0)] = N \eta(i \omega,0) \Big( \chi(i \omega,0) - J \Big) , \\ && \mathcal{S}_{Bulk}[\eta(i \omega,z), \chi(i \omega,z)] = N \int_{0}^{z_{f}} d z \sum_{i \omega} \eta(i \omega,z) \Big( \frac{\partial \chi(i \omega,z) }{\partial z} + \chi(i \omega,z) - f[\chi(i \omega,z)] \Big) , \\ && \mathcal{S}_{IR}[\chi(i \omega,z_{f})] = - \frac{1}{2} \sum_{k} \sum_{i \omega} \ln \Big\{ (- i \omega)^{2} - \Big( 2 \chi(i \omega,z_{f}) \gamma_{k} - J \lambda \Big)^{2} - \Big( 2 \chi(i \omega,z_{f}) \varphi_{k} \Big)^{2} \Big\} . \eqa
\end{widetext}
Here, $\gamma_{k} = \cos k$ denotes a conventional kinetic-energy term and $\varphi_{k} = \sin k$ represents a $p-$wave pairing term in the IR effective action. Now, original fermion degrees of freedom disappear in this effective bosonic action. Instead, they form particle-hole and particle-particle composite fields, which appear in this effective action as bosonic collective excitations given by the order parameter field $\eta(i \omega,z)$ and its canonical conjugate pair, the coupling function $\chi(i \omega,z)$.

\section{Hamilton-Jacobi formulation and emergent metric tensor}

Based on the effective field theory of the previous section, we extract out a metric tensor describing an emergent spacetime with an extra dimension.
%
%
An effective free energy from Eq. (\ref{Kitaev_Model_Emergent_Holography}) should not depend on the IR cutoff $z_{f}$, described by
\bqa && \frac{d}{d z_{f}} \ln Z = 0 . \eqa
This constraint gives rise to
\bqa && 0 = \sum_{i \omega} \Big\{ - \eta(i \omega,z_{f}) \partial_{z_{f}} \chi(i \omega,z_{f}) \nn && + \partial_{z_{f}} \chi(i \omega,z_{f}) \Big\langle \frac{1}{N} \sum_{i = 1}^{N} \psi_{i}^{\dagger}(i \omega) ( \tau_{3} - i \tau_{2} ) \psi_{i+1}(i \omega) \Big\rangle \Big\} , \nn \label{RG_EQ} \eqa
where $\partial_{z_{f}}$ means a derivative with respect to $z$ at $z=z_f$. This equation must be trivial since it is nothing but the definition of an order parameter field $\eta(i \omega,z_{f})$. Performing the path integral of $\int D \chi^{(f)}$ in Eq. (\ref{Holography_Discrete}), we obtain essentially the same equation in a discrete version along the direction of the extra dimension. Actually, this triviality implies self-consistency of our emergent geometric description.

A nontrivial and important point is that Eq. (\ref{RG_EQ}) is the Callan-Symanzik equation for the effective free energy, given by
\bqa && \sum_{i \omega} \Big\{ \gamma^{00} T_{00} + \gamma^{11} T_{11} + \beta_{\chi} \Big\langle \mathcal{O}_{\chi} \Big\rangle \Big\} = 0 . \label{Callan_Symanzik_EQ} \eqa
Here, $\gamma^{00}$ and $\gamma^{11}$ are time and space components of the metric tensor, respectively, and $T_{00}$ and $T_{11}$ are time and space components of the energy-momentum tensor, respectively. $\beta_{\chi}$ is the RG $\beta-$function to describe the evolution of a coupling constant as a function of an energy scale $z_{f}$, and $\Big\langle \mathcal{O}_{\chi} \Big\rangle$ is an expectation value of an observable $\mathcal{O}_{\chi}$, identified with an order parameter, where these variables form a canonical conjugate pair. Comparing Eq. (\ref{RG_EQ}) with Eq. (\ref{Callan_Symanzik_EQ}), we obtain
\bqa && \gamma^{00} T_{00} + \gamma^{11} T_{11} \equiv - \eta(i \omega,z_{f}) \partial_{z_{f}} \chi(i \omega,z_{f}) , \label{RG_Energy_Momentum_Tensor} \\ && \beta_{\chi} \equiv \partial_{z_{f}} \chi(i \omega,z_{f}) = - \chi(i \omega,z_{f}) + f[\chi(i \omega,z_{f})] , \nn \label{RG_Beta_Function} \\ && \Big\langle \mathcal{O}_{\chi} \Big\rangle \equiv \Big\langle \frac{1}{N} \sum_{i = 1}^{N} \psi_{i}^{\dagger}(i \omega) ( \tau_{3} - i \tau_{2} ) \psi_{i+1}(i \omega) \Big\rangle \nn && = \eta(i \omega,z_{f}) . \label{RG_Order_Parameter} \eqa

%
%

It is straightforward to find the energy-momentum tensor, given by
\bqa && T_{00} = - \frac{1}{N} \sum_{i = 1}^{N} \Bigl\{ J \lambda \psi_{i}^{\dagger}(i \omega) \tau_{3} \psi_{i}(i \omega) \nn && - \chi(i \omega,z_{f}) \psi_{i}^{\dagger}(i \omega) ( \tau_{3} - i \tau_{2} ) \psi_{i+1}(i \omega) \Big\} , \\ && T_{11} = \frac{1}{N} \sum_{i = 1}^{N} i \chi(i \omega,z_{f}) \psi_{i}^{\dagger}(i \omega) \tau_{2} \psi_{i+1}(i \omega) . \eqa
This expression is a lattice version of a continuum field theory. Inserting these components into Eq. (\ref{RG_EQ}) with Eq. (\ref{RG_Energy_Momentum_Tensor}), we obtain coupled equations for the emergent metric tensor
\bqa && \chi(i \omega,z_{f}) \Big( \gamma^{00} - \gamma^{11} \Big) = \beta_{\chi} , \\ && \Big( J \lambda - \chi(i \omega,z_{f}) \Big) \gamma^{00} = - \beta_{\chi} . \eqa
Solving these equations, we obtain
\bqa && \gamma^{00} = \frac{(\lambda - 2) e^{z_{f}}}{[2 + (\lambda - 2) e^{z_{f}}][1 + (\lambda - 2) e^{z_{f}}]} , \\ && \gamma^{11} = \frac{(\lambda - 2) e^{z_{f}}}{ 1 + (\lambda - 2) e^{z_{f}} } , \eqa
where the low-frequency limit has been taken into account. This approximation will be justified in the discussion of the holographic entanglement entropy.
Under the following redefinitions
\bqa && z \longrightarrow 2 z , ~~~~~ \tau \longrightarrow \sqrt{\lambda - 2} \tau , ~~~~~ x \longrightarrow \sqrt{\frac{\lambda - 2}{2}} x , \nonumber \eqa
we find an emergent metric structure
\bqa && d s^{2} = d z^{2} + g_{00} d \tau^{2} + g_{11} d x^{2} \eqa
in the normal coordinate system, where the time and space components are
\bqa && g_{00} = \frac{[2 + (\lambda - 2) e^{2 z}][1 + (\lambda - 2) e^{2 z}]}{2 e^{2 z}} , \label{Metric_00} \\ && g_{11} = \frac{ 1 + (\lambda - 2) e^{2 z} }{e^{2 z}} , \label{Metric_11} \eqa
respectively. We note that the $z \rightarrow - \infty$ limit gives rise to an AdS$_{3}$ metric, identified with an insulating UV fixed point in our real-space RG construction.

{At the quantum critical point ($\lambda = 2$), the emergent spacetime is exactly described by the AdS$_{3}$ metric $d s^{2} = d z^{2} + e^{- 2 z} (d \tau^{2} + d x^{2})$. On the other hand, the metric of the topologically trivial superconducting phase ($\lambda > 2$) is given by $d s^{2} = d z^{2} + \frac{ (\lambda - 2)^{2} }{2} e^{2 z} d \tau^{2} + (\lambda - 2) d x^{2}$ in the IR limit ($z \longrightarrow \infty$). Intriguingly, it describes to the metric of AdS$_2 \times R$.} In the topological superconducting phase with $1 \leq \lambda < 2$, the extra dimension ends at $z_{c} = \frac{1}{2} \ln \Big( \frac{1}{2 - \lambda} \Big)$, where $g_{00}(z_{c}) = 0$ and $g_{11}(z_{c}) = 0$. In other words, $z_{c}$ plays the role of an IR cutoff. The metric is given by $g_{00}(0 \leq z < z_{c}) = \frac{[2 + (\lambda - 2) e^{2 z}][1 + (\lambda - 2) e^{2 z}]}{2 e^{2 z}}$ and $g_{11}(0 \leq z < z_{c}) = \frac{ 1 + (\lambda - 2) e^{2 z} }{e^{2 z}}$. This result is consistent with that of Ref. \cite{From_Field_Theory_To_Emergent_Gravity}. In Ref. \cite{From_Field_Theory_To_Emergent_Gravity}, although the IR cutoff is different from the black hole horizon,  it was called ``{\it horizon}". The emergence of such a horizon has been proposed to be a fingerprint of a quantum phase transition in the geometric description of Ref. \cite{From_Field_Theory_To_Emergent_Gravity}. Following Ref. \cite{From_Field_Theory_To_Emergent_Gravity}, from now on, we also utilize the horizon in order to indicate the IR cutoff. In the case of $0 \leq \lambda < 1$ the IR cutoff $z_{c} = \frac{1}{2} \ln \Big( \frac{1}{2 - \lambda} \Big)$ becomes negative. This implies $g_{00}(z_{c} < 0 \leq z) < 0$ and $g_{11}(z_{c} < 0 \leq z) < 0$. The emergent metric is not well-defined in $0 \leq \lambda < 1$. However, we emphasize that this parameter region is not special. The reason why the metric is not well defined is that the UV boundary starts from $z = 0$. If we start from $z_{i} < 0$, the metric would be well defined in $z_{i} < z < z_{c} < 0$. Of course, we choose $z_{i} < z_{c}$.

\begin{figure}[h]
\centering
\includegraphics[scale=0.45]{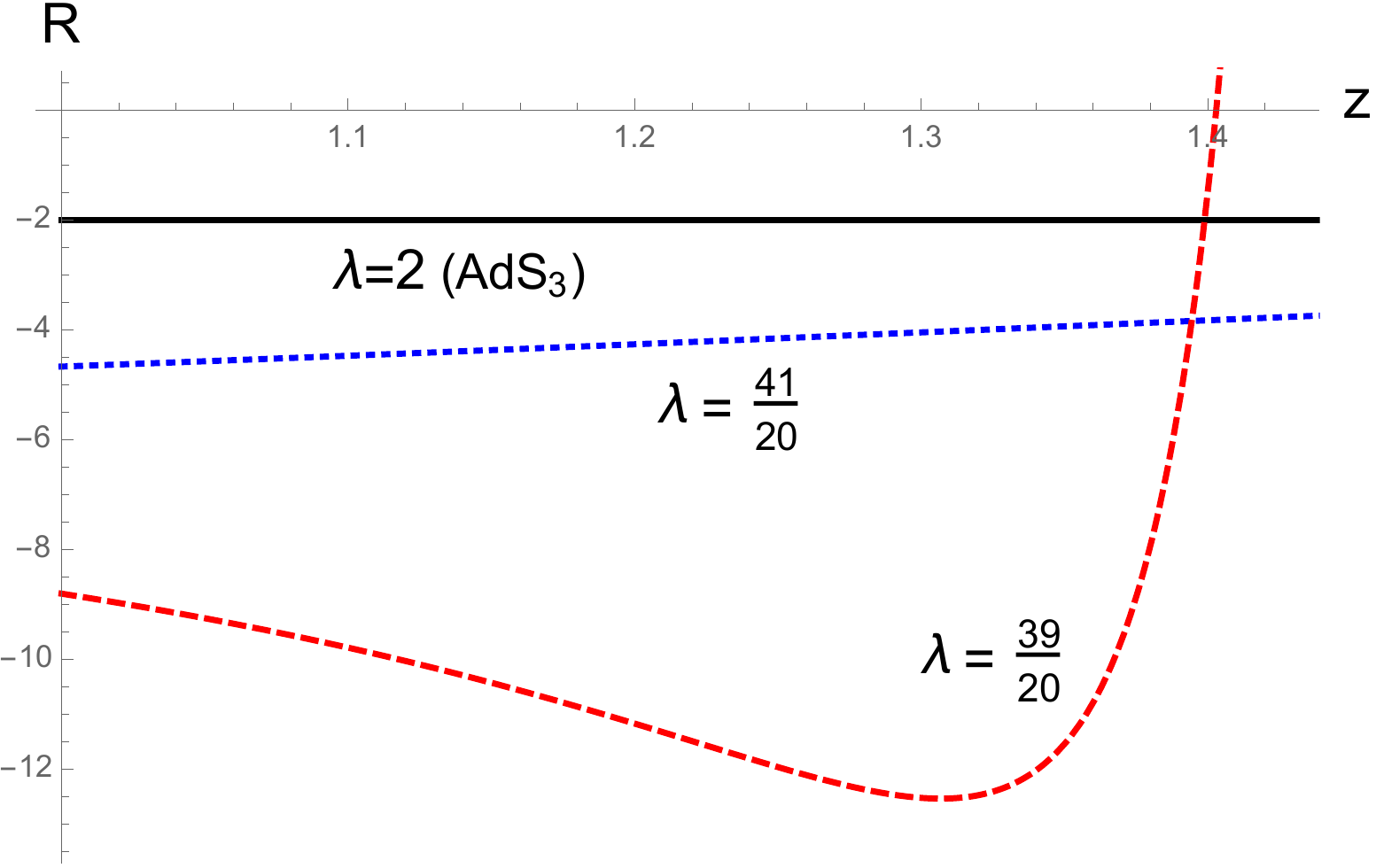}
\caption{Emergent Ricci curvature given by Eqs. (\ref{Metric_00}) and (\ref{Metric_11}) for the quantum critical point ($\lambda = 2.0$), the topologically trivial ($\lambda = 2.05$), and nontrivial ($\lambda = 1.95$) superconducting phases. We emphasize that the Ricci curvature diverges at $z = z_{c}$ in the topological superconducting phase, which may be identified with a horizon. Emergence of such a horizon in a dense phase is consistent with a recent study \cite{From_Field_Theory_To_Emergent_Gravity}, where the existence of the horizon is a fingerprint of a quantum phase transition.}
\label{Emergent_Metric_Horizon}
\end{figure}

Fig. \ref{Emergent_Metric_Horizon} shows Ricci curvatures as a function of the extra dimension for the topological superconducting phase, the quantum critical point, and the normal superconducting state, respectively. The UV AdS$_{3}$ ($z \rightarrow - \infty$) does not evolve at the quantum critical point, which still remains to be AdS$_{3}$ at IR. On the other hand, the UV AdS$_{3}$ develops a horizon at $z_{c}$ in the topological superconducting phase, where the Ricci curvature diverges. In the normal superconducting state, the curvature again converges on a negative constant at IR . This is because a new  AdS$_2 \times R$ metric occurs in the IR limit, as mentioned before.

\section{Entanglement entropy}

In order to clarify the physical meaning of this metric structure, we investigate the holographic entanglement entropy \cite{Ryu_Takayanagi_Formula}, given by the minimal surface area along the emergent spacetime direction,
\bqa && S_{E} = \frac{1}{4 G} \int_{-l/2}^{l/2} d x \sqrt{g_{11}[z(x)] + \Big( \frac{d z(x)}{d x} \Big)^{2}} , \eqa
where the subsystem size is $l$. Here, $G$ is the Newton's constant in this three dimensional spacetime. It is straightforward to reformulate this expression as follows \cite{Holographic_Entanglement_Entropy_CYP}
\bqa && S_{E} (z_0) = \frac{1}{2 G} \int_0^{z_0} dz \sqrt{\frac{ g_{11} }{ g_{11} - g^{0}_{11}} } , \eqa
where $z_{0} = z(0)$ is a maximum value in the extra dimension, given by the turning-point condition $\frac{d z(x)}{d x} \Big|_{x = 0} = 0$ and determined by the subsystem size
\bqa && l = 2 \int_0^{z_0} dz \sqrt{\frac{g^{0}_{11}}{g_{11} (g_{11} - g^{0}_{11})}} . \eqa
Here, $g^{0}_{11} = g _{11}(z_{0})$ is the space-component metric at the turning point.

In order to discuss the IR behavior of the holographic entanglement entropy, it is convenient to consider $\lambda \longrightarrow \lambda + 1$. Then, the quantum critical point shifts from $\lambda_{c} = 2$ to $\lambda_{c} = 1$. At the quantum critical point described by the AdS$_{3}$ metric of $g_{00} = g_{11} = e^{-2 z}$, we find the entanglement entropy
\bqa && S_{E} = \frac{1}{2 G} \ln \left( \frac{\sqrt{4+l^2}}{2} + \frac{l}{2} \right) , \eqa
where the maximum value of the extra dimension is $z_0 = \ln \frac{\sqrt{4+l^2}}{2}$. This expression becomes reduced into $S_{E} \approx \frac{1}{2 G} \ln l$ in the $l \rightarrow \infty$ limit. This logarithmically divergent behavior is consistent with that of a conformal field theory \cite{Cardy_CFT}. On the other hand, the metric $g_{11}$ has a root at $z_c = \frac{1}{2} \ln \frac{1}{1-\lambda}$ in the case of $0 \leq \lambda < 1$. The metric above this critical value ($z > z_{c}$) becomes negative, so the radial direction is restricted to $0 \leq z \leq z_c$. In the large $l$ limit where $z_c \ll z_0$, the leading contribution to the entanglement entropy comes from
\bqa S_{E} = \frac{1}{4 G} \left( \ln \frac{(1 +\sqrt{\lambda})^2}{1-\lambda} - 2 \sqrt{\lambda} \right) . \eqa
Near the critical point ($\lambda \approx 1$) where the central charge of the conformal field theory is related to the Newton constant of the dual gravity as $c = \frac{3R}{2G}$, the entanglement entropy shows a logarithmic scaling behavior
\bqa && S_{E} \approx \frac{c}{6} \ln \frac{1}{1-\lambda} . \eqa
Here, we set $R=1$ and $c=N_{f}/2$, where $N_{f}$ is the number of fermion flavors. Intriguingly, this result is perfectly matched to Cardy's result for $N_{f}=1$, where the correlation length is given by $\xi = (1-\lambda)^{-1}$ \cite{Cardy_CFT}.

\begin{figure}[h]
\centering
\includegraphics[scale=0.37]{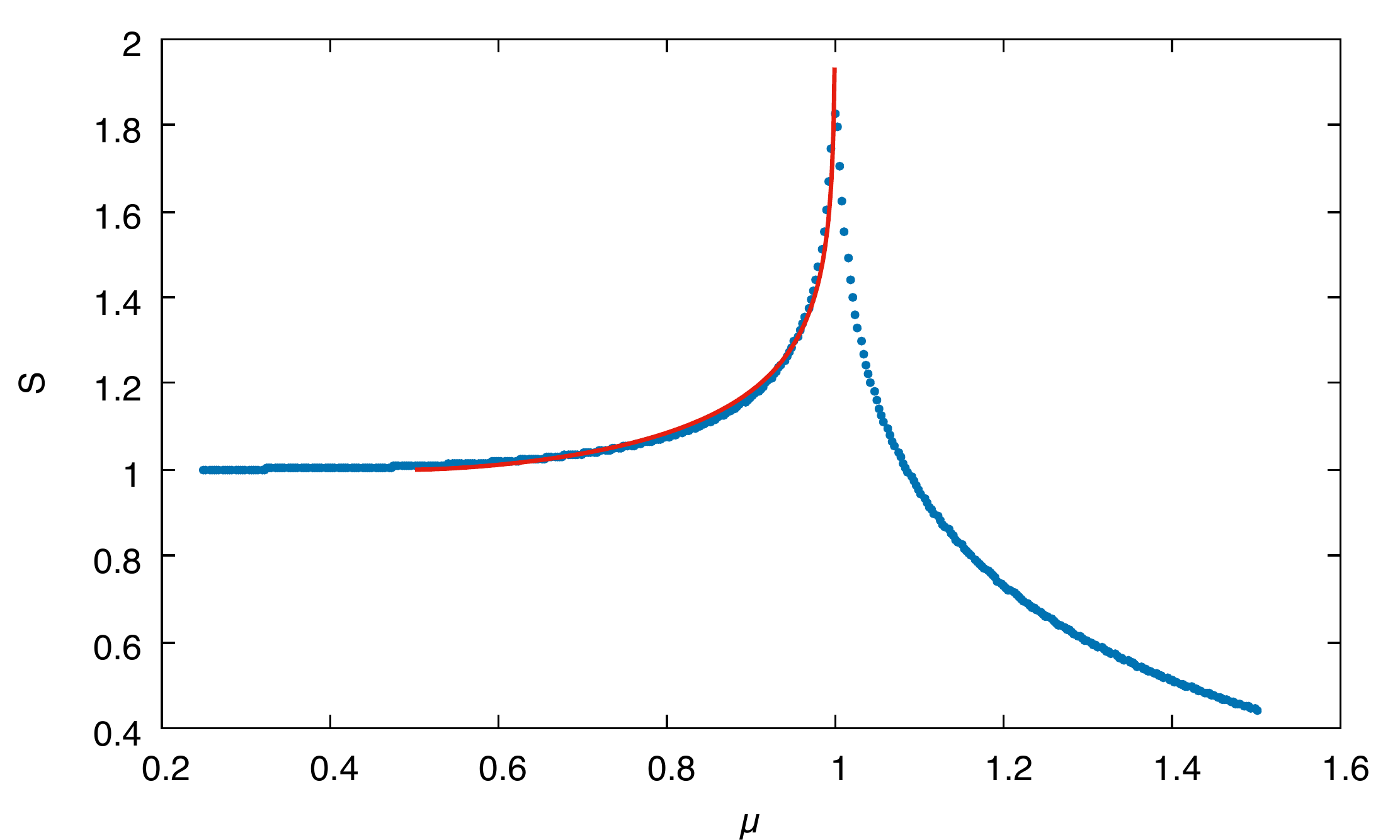}
\caption{Comparison between the Ryu-Takayanagi Formula based on the emergent metric [Eqs. (\ref{Metric_00}) and (\ref{Metric_11})] and the entanglement entropy based on the lattice model [Eq. (\ref{Kitaev_Model})]. The Ryu-Takayanagi Formula describes the red-thick line and the field-theory entanglement entropy represents blue-dots. Here, we have $\mu = \frac{\lambda + 1}{2}$. The total number of lattice sites is $L = 400$. See Ref. \cite{EE} for how to calculate the entanglement entropy from the lattice model directly. It is surprising for both quantities to match almost perfectly not only near the quantum critical point but also far away from it in the topological superconducting phase. We note that a constant value of $log_{2} 2$ is added into the Ryu-Takayanagi Formula, where our renormalization group procedure does not take into account effects of boundary Majorana-fermion zero modes. This comparison leads us to suspect that the Ryu-Takayanagi Formula based on the emergent metric [Eqs. (\ref{Metric_00}) and (\ref{Metric_11})] may be identified with an entanglement entropy although we did not prove that this emergent metric is governed by the Einstein equation.}
\label{Holographic_Entanglement_Entropy}
\end{figure}

Fig. \ref{Holographic_Entanglement_Entropy} shows the comparison between the holographic entanglement entropy based on Eqs. (\ref{Metric_00}) and (\ref{Metric_11}) and the field-theory entanglement entropy of Eq. (\ref{Kitaev_Model}). As proven above, both coincide not only at the quantum critical point but also near it in the topological superconducting phase. However, rather far away from the quantum critical point in the topological superconducting phase, it turns out that they match quite well if a constant value of $log_{2} 2$ is added into the holographic entanglement entropy. This constant value for the entanglement entropy is nothing but counting the number of Majorana-fermion zero-energy states at a boundary. Unfortunately, this boundary-mode contribution is not taken into account in our RG scheme since we considered a periodic boundary condition during the RG procedure. In order to introduce this topological effect into the entanglement entropy within our geometric construction, we need to modify our RG scheme. Except for this aspect, both entanglement entropy show remarkable match.

%
%

\section{Discussion}

One may criticize to identify the Ryu-Takayanagi formula with the entanglement entropy because it is not clear at all whether or not the emergent metric satisfies the Einstein equation. However, it is true that Fig. \ref{Holographic_Entanglement_Entropy} is quite impressive. How can we figure out the governing equation for the emergent metric tensor? The bulk metric corresponds to the running of the coupling of the stress tensor. In order for the RG dependence of this coupling to be determined, the coupling may be introduced in the original model at UV. That is, apart from the coupling $\chi^{(0)}$ dual to the scalar operator, the effective action may also be a function of an arbitrary $(1+1)$ dimensional metric. The RG analysis should then lead to two independent RG flow equations, i.e. expressions for the beta functions, one for the scalar coupling and one for the running $(1+1)$ dimensional metric, i.e. the coupling of the stress tensor. Solving this system of first order equations may then correctly determine the RG flow of the couplings, including the bulk metric.

It would be interesting to apply our geometric construction into strongly coupled quantum field theories, of course. Recently, we could find an effective geometric description for the Kondo effect \cite{Emergent_Holography_Interaction}, where a $[\eta(i \omega,z)]^{2}-$like term appears to give a $\Big( \frac{\partial \chi(i \omega,z) }{\partial z} \Big)^{2}-$like term, which gives rise to a nontrivial evolution of a coupling function. More surprisingly, the $z_{f} \rightarrow 0$ limit in this geometric description reproduced leading $1/N_{f}$ quantum corrections in the slave-boson mean-field theory of the Kondo effect, where $N_{f}$ is spin degeneracy. This shows the role of an emergent extra dimension clearly.

\section*{Acknowledgement}

K.-S. Kim was supported by the Ministry of Education, Science, and Technology (No. NRF-2015R1C1A1A01051629 and No. 2011-0030046) of the National Research Foundation of Korea (NRF) and by TJ Park Science Fellowship of the POSCO TJ Park Foundation. M. Park was supported by Basic Science Research Program through the National Research Foundation of Korea funded by the Ministry of Education (NRF-2016R1D1A1B03933399). J. Cho was supported by the R$\&$D Convergence Program of NST (National Research Council of Science and Technology) of Republic of Korea (Grant No. CAP-15-08-KRISS). C. Park was supported by Basic Science Research Program through the National Research Foundation of Korea funded by the Ministry of Education (NRF-2016R1D1A1B03932371) and also by the Korea Ministry of Education, Science and Technology, Gyeongsangbuk-Do and Pohang City. This work was also supported by the POSTECH Basic Science Research Institute Grant (2016). We would like to appreciate fruitful discussions in the APCTP Focus program ``Lecture Series on Beyond Landau Fermi Liquid and BCS Superconductivity near Quantum Criticality" (2016). K.-S. Kim would like to express his sincere thanks to Dr. G.-Y. Cho for his suggestion on the gravity reformulation for the transverse-field Ising model. We also thank Eoin O. Colgain for helpful discussions.

\end{document}